\documentclass[12pt]{article}

\usepackage{scicite}
\usepackage{bm}
\usepackage{amsmath}
 \usepackage{braket}

\usepackage{times}
\usepackage{graphicx}
\newenvironment{sciabstract}{%
\begin{quote} \bf}
{\end{quote}}

\newcounter{lastnote}

\title{Quantum Spin-Hall Effect of Light at Bound States in the Continuum}

\author
{Gianluigi Zito,$^{1\ast}$ Silvia Romano,$^{2}$  Stefano Cabrini,$^{3}$ \\
  Giuseppe Calafiore,$^{3}$ Anna Chiara De Luca,$^{1}$ Erika Penzo,$^{3}$ Vito Mocella$^{2\ast}$\\
\\
\normalsize{$^{1}$ National Research Council IBP, }\\
\normalsize{Via Pietro Castellino, Naples, 80131, Italy}\\
\normalsize{$^{2}$ National Research Council IMM, }\\
\normalsize{Via Pietro Castellino, Naples, 80131, Italy}\\
\normalsize{$^{3}$ Molecular Foundry, }\\
\normalsize{Berkeley, CA 94720, USA}
\\
\normalsize{$^\ast$To whom correspondence should be addressed; E-mail:  g.zito@ibp.cnr.it, vito.mocella@imm.na.cnr.it}
}

\date{2017}

\begin{document} 


\baselineskip24pt


\maketitle


\begin{sciabstract}
The discovery of the topological nature of free-space light and its quantum chiral behavior
has recently raised large attention. This important scientific endeavor features spin-based integrated quantum technologies. 
Herein, we discuss a novel phenomenon based on a resonantly-enhanced quantum spin-Hall transport of light observed in a dielectric resonator operating near the bound-state-in-continuum (BIC) regime. The BIC mode is characterized by a transverse photonic spin angular momentum density extended on a macroscopic area. As such, the experimental excited mode in near-BIC regime generates resonant surface waves characterized by spin-momentum locking and that propagate along the symmetry axes of the structure. In addition, the generated side waves are interpreted as an abrupt nonparaxial redirection of the exciting far field light, which is responsible for the spin-to-orbital angular momentum conversion evidenced in the spin-orbit asymmetry measured in the intensity of the side waves. The experimental results are in excellent agreement with a model that combines geometric parallel transport of light polarization and spin-momentum locking. In addition, breaking the excitation symmetry leads to a total spin-directive coupling. Our results reveal the possibility of a BIC-enhanced macroscopic spin-directive coupling, a novel fundamental mechanism of light-spin manipulation that will have strong impact on emerging quantum technologies.
%This all-dielectric approach provides excellent efficiency. Its robustness and easy integrability and scalability are promising for an effective application to quantum communication protocols and multiplatform  photonic device implementation.  
\end{sciabstract}

% In setting up this template for *Science* papers, we've used both
% the \section* command and the \paragraph* command for topical
% divisions.  Which you use will of course depend on the type of paper
% you're writing.  Review Articles tend to have displayed headings, for
% which \section* is more appropriate; Research Articles, when they have
% formal topical divisions at all, tend to signal them with bold text
% that runs into the paragraph, for which \paragraph* is the right
% choice.  Either way, use the asterisk (*) modifier, as shown, to

\paragraph*{Introduction.}
 
In solid state physics, the quantum spin Hall effect (QSHE) is characterized by unidirectional edge spin transport of electrons \cite{kane2005}. Edge states are chiral, \textit{i.e.} opposite spin states propagate in opposite directions without dissipation and are protected by time reversal symmetry. This mechanism is termed spin-momentum locking and such topological transport states gave rise to a new class of materials termed topological insulators \cite{qi2011}. 
 %\cite{bliokh2008,bliokh2015,liu2017,berry1987,andrews2012angular} 
Recently, the topological nature of free-space light and its quantum chiral behavior have raised large attention  \cite{bliokh2015quantum, bliokh2014extraordinary, lodahl2016chiral}. Indeeed, unusual transverse spin angular momentum (SAM) density arising in inhomogeneous optical fields is the hallmark of an intrinsic quantum spin Hall effect exhibited by free-space light \cite{bliokh2015}.  As a result of transverse SAM, evanescent waves with opposite spins at the interface between two media travel in opposite directions obeying spin-momentum locking. This is a universal property of nanoscale waveguiding mechanisms  and occurs in evanescent structured waves in plasmonic systems, photonic crystal waveguides and whispering-gallery-mode resonators \cite{aiello2015,van2016universal}.  
The resulting spin-directive coupling paves the way towards novel quantum communication protocols \cite{le2015,mitsch2014,sollner2015}.  Photonic metasurfaces, \textit{i.e.} structures with a reduced dimensionality, are excellent candidates for controlling and devise spin-orbit coupling processes in integrated systems \cite{liu2017,maguid2016photonic,khanikaev2013photonic,yin2013photonic,shitrit2013spin,lin2014dielectric,yu2011light}. Major results have been demonstrated in nanoscale dielectric structures with near-field coupling \cite{petersen2014chiral,sollner2015} and in plasmonic systems \cite{huang2013helicity,Zayats2014} . 
However, surface plasmon polariton waves in metal metasurfaces are quickly extinguished by metal absorption and scattering losses.  

Herein, we report on the first  experimental observation of a  \textit{macroscopic} quantum spin Hall effect of light with an all-dielectric photonic crystal metasurface (PhCM) resonator designed to support a bound state in the continuum (BIC)  at optical frequencies \cite{hsu2013,lu2014topological,mocella2015giant}. This resonator operating in near-BIC regime represents a leap forward for controlling quantum spin-Hall directive coupling of light. The concept of BIC was introduced by von Neumann and Wigner
in quantum mechanics as unusual spatially localized states of electron waves \cite{von1929} but BICs occurr in all wave phenomena \cite{friedrich1985}. 
In particular, \textit{optical} bound states in the continuum have been realized in photonic crystal thin slabs where `guided resonances' may have ideally infinite
lifetimes at the BIC \cite{hsu2013,fan2002analysis}. This occurs at specific conditions of field symmetry mismatch or topological constraints with respect to out-of-plane propagating waves in free-space \cite{sakoda2004,zhen2014topological,lu2014topological}. Recently, supercavity lasing in the BIC regime \cite{kodigala2017} and polarization BICs \cite{gomis2017} have been demonstrated. However,  spin-momentum locking at the BIC  - signature of the QSHE effect -  so far has never been reported. Herein, we observe a spin-momentum locking in near-BIC regime with an efficiency that reaches $\sim$ 100$\%$ when breaking the excitation symmetry. This novel phenomenon is correlated with the transverse SAM of the structured \textit{evanescent} BIC field. In addition,  collimated surface waves with well-defined linear momenta  imposed by the PhC symmetry are observed and outcoupled from the the system. The abrupt change of photon trajectories  with respect to the normal incidence excitation is at origin of  nonparaxial spin-to-orbital angular momentum conversion (SOC) \cite{bliokh2011,bliokh2015,bliokh2015transverse}. Indeed, the redirected waves show a nontrivial orbital angular momentum (OAM) \cite{allen2003optical,marrucci2006optical, andrews2012angular}.
%Indeed, incident photons are redirected in the plane of the PhC, \textit{i.e.} at a $\pi/2$-angle with respect to the normal incidence. 
Our numerical simulations predict the spin-momentum locking in agreement with the existence of a transverse SAM at the BIC. A model based on geometric parallel transport of light polarization \cite{bliokh2011,bekshaev2011internal} allows us to interpret with excellent agreement the spin-dependent intensity and polarization of the outcoupled radiation.  These results reveal a novel mechanism of light manipulation that is based on a spin-directive coupling resonantly enhanced by a bound state in the continuum, and point out new exciting opportunities for spin-directional control of surface light waves, manipulation of orbital degrees of freedom, and potential scalability in integrated photonic circuits for chiral quantum technology applications \cite{petersen2014chiral,sollner2015,liu2017,maguid2016photonic,
 khanikaev2013photonic,yin2013photonic,shitrit2013spin,lin2014dielectric,yu2011light}. \\
 
 \paragraph*{Results.} 
 
We studied a dielectric geometry consisting of a square lattice of cylindrical air holes etched in a thin film (membrane) of silicon nitride (Si$_3$N$_4$). The resonator is  designed to be sandwiched between air (top side) and a supporting quartz (SiO$_2$) coverslip (Fig. 1) and is transparent in both visible and infrared ranges of light frequencies.
 We fabricated several dielectric PhC metasurfaces with tuned etching parameters in order to maximize the experimental BIC resonance properties (Methods). Design parameters were optimized to have a BIC at visible frequencies, and, alternatively, at infrared frequencies in order to demonstrate that this phenomenon is strictly related to the designed BIC structure and can be totally independent from the BIC wavelength. Let us first introduce the physical phenomenon expected to occur at the BIC. 
 
We carried out extensive numerical simulations of the resonator near the BIC regime. We recognized that - as expected - the confinement imposed by a BIC generates surface waves whose transverse SAM mediates the spin-momentum locking. Numerical modeling optimization was carried out using a rigorous coupled wave approach (in-house developed), finite element method and finite-difference time domain (FDTD) simulations (commercial platforms).  Of particular relevance is the numerical analysis carried out with the full three-dimensional FDTD algorithm. This revealed the intriguing behavior of the modes that can be excited at the BIC frequency for right-circularly polarized (RCP) and left-circularly polarized (LCP) input plane waves at normal incidence (Fig. 2).
At normal incidence ($\bm{\Gamma}$-point) of the Brillouin zone, a square-lattice PhC has a C4$\nu$ point group symmetry. Indeed, for $\lambda$ close to the first Bragg condition  $ a \sim  \lambda/n_{e}$ (with $n_{e}$ effective refractive index), 4 dominant guided modes exist for each polarization in correspondence of the $(\pm1, 0)$ and $(0, \pm1)$ lattice points and are characterized by high quality factor \cite{mocella2015giant}. Our simulations reveal that the optical mode is consistent with a singly-degenerate resonance \cite{fan2002analysis} having the symmetries of the crystal (Fig. 2a,b).
%and approaching the infinite theoretical intrinsic quality factor ($Q_{V} > 10^{8}$). Of course, the experimental total $Q$-factor is limited by the finite area of the crystal and scattering losses \cite{mocella2015giant}.
This symmetry-protected BIC occurs at a free-space wavelength $\lambda = 540 $ nm for the lattice parameters given in Methods. A crucial point is that the thickness $h$ of the resonant membrane results $\sim \lambda/10$ - so we can refer to this resonator as a metasurface.  In Fig. 2a-b, the electric field intensity $|E|^2$ in the PhC unit cell is represented with the superimposed maps of the electric field $\bm{E}$ in the $(x, y)$-plane of the membrane, for both  RCP and LCP incident radiation, respectively. The electric field structure ($\bm{E}$) appears as a vortex-antivortex pattern in the $(x, y)$-plane (Fig. 2a,b). In the $z$-direction, the field intensity $|E|^2$ has the clear characteristics of a confined surface wave, strongly enhanced at the interface  with the SiO$_2$. 
Indeed, the frequency for which the PhCM is designed to support a BIC is $\omega_{n}=a/\lambda=0.667$  ($\Gamma$ point), thus below the light line of the quartz support ($n$ = 1.46, $a/\lambda=1/n=0.685$)  and also far from the light line in air ($a/\lambda=1$). 
The wavevectors in the PhC have modulus $k = k_{tangential} =2 \pi/a$, are parallel to the $(x, y)$-plane and are
conserved across both interfaces with air and quartz substrate. Along these surfaces, the wave cannot couple to the far field having
a purely imaginary $z$-component, $k_{z}=\sqrt{k_{air,SiO_{2}}^{2}-k_{tangential}^{2}} \doteq i \kappa_{1,2}$. Indeed, the tangential component is larger than
the wavevector modulus in both SiO$_{2}$ where $ k_{SiO_{2}}=2\pi/a>2\pi n/\lambda $ and air for which $k_{air} = 2\pi/\lambda$.  Therefore, the guided resonance is a Bloch wave exponentially decaying in both $\pm \hat{\bm{z}}$, that is $\sim \rm{e}^{-\kappa_{1} (z - h)}$ in air and $\sim \rm{e}^{-\kappa_{2} |z|}$ in SiO$_2$ ($\kappa_1 > \kappa_2$).   
 
The $\bm{E}$-field shows odd mirror symmetry with respect to the $z$-axis for opposite exciting spin states and evolves with cycloidal rotations in the meridional planes close to the membrane (Fig. 2a,b), consistently with the existence of transverse SAM.  Concurrently, the 3$D$ time averaged Poynting vector density $\bm{\Pi} = (1/2){\rm Re}\left[\bm{E^\star} \times \bm{H}\right]
$ shows nearly perfect light confinement in the $(x, y)$-plane of the PhC, with reversed vectors orientation from RCP to LCP (Fig. 2c,d). 
We calculated the SAM density $\bm{S} ={\rm Im}\left[\epsilon_{0}\bm{E^{\star}} \times \bm{E}+\mu_{0}\bm{H^{\star}} \times\bm{H}\right]/4\omega$, with $\bm{H}$ magnetic field and $\omega$ angular frequency of the  radiation, associated to this BIC mode eliciting its transverse components $\bm{S}_{\perp}$ with respect to the propagation directions confined in the $(x,y)$-plane (Methods). Figure 2e shows a non-null electromagnetic SAM density along the $z$-axis - hence necessarily transverse with respect to $\bm{\Pi}$ - when the BIC resonance is excited. The SAM has always components orthogonal to the linear momentum density in each point of the unit cell. Indeed, the SAM is orthogonal to $\bm{\Pi}$ in such a way that it may lie in the $(x,y)$-plane with a nearly zero $z$-component, or it may have a dominant $z$-component pointing out of the crystal plane. The specific direction of the transverse spin depends on the position. The net average transverse spin is not zero and has a component along the $z$-direction, thus transverse to the Poynting vector.  In particular, inspecting the dashed area, the detail of the arrow maps of   $\bm{\Pi}$ and $\bm{S}$ for RCP and LCP reveals clearly that the direction of propagation is locked to the spin direction. Depending on the helicity of the excitation, the transverse $\bm{S}$ component to which the input spin is coupled determines the orientation of the momentum $\bm{\Pi}$, enabling the spin-directive coupling. In brief, the SAM density has always a  component  perpendicular to the direction of propagation, with orientation locked to sign of the linear momentum, whether such SAM component lies in the $(x, y)$-plane or is directed along the $z$-axis. These numerical results consistently interpret our experimental results, as discussed in the following.

State-of-the-art fabrication methodologies allowed us to produce large-area  PhC membranes (1 mm$^2$) of Si$_3$N$_4$ with thickness down to  $h = 54$ nm for a BIC mode expected at $\sim$ 540 nm (Fig. 1). Si$_3$N$_4$ thin films were deposited on commercial SiO$_{2}$ coverslips of 120 $\mu$m of thickness (details on modeling and fabrication can be found in Methods).  A schematic layout of the setup for the optical characterization of the PhCM is shown in Fig. 3a. The collected transmitted spectrum was characterized with a supercontinuum laser source. A representative spectrum collected for the PhCM working in the visible range is shown in Fig. 3b. At normal incidence $\theta_{i} = 0^{\circ}$ the spectrum did not show any dependence on the input polarization state. It exhibited two spectral dips at $\lambda_{1}$ = 541.4 nm and $\lambda_{2}$ = 550.0 nm, corresponding respectively to a BIC mode (guided resonance) and to a guided mode. 
It is worth mentioning that although the intrinsic quality factor ($Q_{V}$) of a BIC peak tends to infinity, the resonance linewidth measurable with a far-field approach
 is limited by finite sample size ($Q_{H}$), effective collimation and scattering losses due to the sample imperfections ($Q_{A}$)  (Methods) \cite{gansch2016measurement}. 
We measured a total $Q\sim 10^3$ corresponding to a linewidth $\simeq$ 0.5 nm (Fig. 3b) as due to finite lateral confinement in agreement with previous reports \cite{gansch2016measurement,hsu2013}. 

Remarkably, when the light source coupled with the bound state at $\lambda_{1}$, light trapped inside the membrane (at the center of the supporting glass in correspondance of the green laser spot of 0.8 mm of diameter) generated four coherent collimated beams that were observed along the four-fold symmetry directions of the PhC, with well-defined linear momenta $\textbf{k}_{R} = -k \bm{\hat{y}}, \textbf{k}_{L}=+k\bm{\hat{y}}, \textbf{k}_{T}=+k\bm{\hat{x}},\textbf{k}_{B}=-k\bm{\hat{x}}$ as experimentally shown in Fig. 3c (see also the scheme in Fig. 3a).    We used an acousto-optical tunable filter to tune the incident wavelength to $\lambda_{1}$ or $\lambda_{2}$ - with linewidth $\sim 4$ nm.  At $\lambda_1$, light transmitted through the sample decreased down to 20$\%$ of its peak incident intensity $I_{\rm in}$ within the excitation spectral window, showing a clear dip with 0.5-nm width (Fig. S1). As above mentioned, most of this power was efficiently conveyed to side waves, with up to $0.1 I_{\rm in}$ per wave typically detected in the far field on the sides of the sample as discussed later.
Conversely, at $\lambda_2$, despite the excitation of a guided mode of the crystal, no light was visible around  the PhCM   (Fig. 3d). The collected beam intensity on the lateral sides was of the order of the noise level. In other words, the redirection effect was observable in the resonator only close enough to the BIC regime $\lambda_1$. 

Let us discuss the origin of the side waves observed at the BIC.  The surface wave at the air/Si$_3$N$_4$ boundary is more tightly confined and weaker than the wave at the Si$_3$N$_4$/SiO$_{2}$ interface. The latter is favorably coupled  to the surface wave propagating at the interface Si$_3$N$_4$/SiO$_2$ (Si$_3$N$_4$ covers all quartz area) and partly leaks to the quartz substrate allowing its propagation and finally detection in the far field. Therefore, the side waves stem from the evanescent surface waves generated at the boundary Si$_3$N$_4/$SiO$_{2}$. The asymmetric PhC slab configuration helped detecting them because guided by the supporting SiO$_2$, which allowed outcoupling the side beams from the system (Fig. S2). 

Of course, the linear momentum of these waves is parallel to the $(x, y)$-plane of the PhCM and directed along the lattice axes as imposed by the system symmetry. The evanescent field is then characterized by a transverse SAM (Fig. 2). As a consequence, a chiral behavior is expected to occur in the intensity of such waves. Figure 3e depicts the principle of the QSHE at the BIC for the waves guided in SiO$_2$ in the four-fold symmetric PhCM.
Indeed, we observed a dependence on the input polarization state in the intensity of the outcoupled redirected side beams. The layout of the experiment is shown in more detail in Fig. 4a. 
Tunable waveplate retarders were used to set the input polarization $\bm{E_i}$  with a polarization-state generator (PSG) consisting of a polarizer (axis // $\bm{\hat{x}}$), a half wave plate (HWP) and a quarter wave plate (QWP). 

The experimental Stokes polarimetry carried out on the redirected beams revealed a significant degree of circular polarization, $|s_{3}|> 60\%$, with opposite signs of $s_3$ on opposite sides of the sample, (R, L) and (T, B), respectively, regardless of the input polarization state incident on the sample.
 
The intensity $I(\alpha,\beta)$  of the redirected beams in the far field was measured as a function of the angle $\beta$ for fixed values of $\alpha$. In particular, we fixed the HWP angle $\alpha$ to two values,  $\alpha^{S} = 0^{\circ}$ ($S$-polarized input on the QWP) and $\alpha^{P} = 45^{\circ}$ ($P$-polarized input on the QWP), whereas $\beta \in (0, 360^{\circ})$. This trick allows to reverse the sign of the circular polarization generated by the QWP so to compare the overall behavior of the intensity for a fixed generated wave on the sides of the sample. As for instance, comparing the curves  for the bottom side-waves ($\bm{k}_{B}$) in Fig. 4c obtained for $\alpha^{S}$ (left) and $\alpha^{P}$ (right), we observe that for $\beta = 45^{\circ}$ - for which the input state is RCP in the left panel where it is LCP in the right panel - the intensity measured is maximum for the LCP case. In other words, LCP photons are preferentially detected in the bottom redirected beam $\bm{k}_{B}$. The overall behavior as a function of $\beta$ is then reversed. This is the first experimental hallmark of a QSHE of light. In analogy to the case reported in \cite{petersen2014chiral} for a 2-fold symmetric nanofiber and symmetric excitation, it is worth noticing that also in our case symmetry breaking is not necessary to observe QSHE. Indeed, it is the strong spin-orbit interaction to generate the asymmetry of the side wave intensities. As for the case reported in \cite{petersen2014chiral} where the exciting incident spin coupled to the transverse spin of the nanofiber modes (parallel to the incident spin), our incident SAM, parallel to $\bm{\hat{z}}$, locks to the photon flow mediated by the $z-$component of the transverse SAM of the BIC mode, with opposite propagation directions in the $(x, y)$-plane depending on the sign of the spin of the incident optical field. This explains why the side waves  experimentally manifest a macroscopic chiral behavior, which could not be explained without spin-momentum locking (spin-orbit interaction) into a C4$\nu$ symmetric PhC. The macroscopic nature of the effect is associated to the resonantly-enhanced routing provided by the BIC. Furthermore, the actual projection of the incident photons to redirected side waves, from $\bm{\hat{z}}$ to the $(x, y)$-plane of the PhC ($\theta_{r}=\pi/2$, Fig. 3a,c), can be associated to another concurrent phenomenon, a spin-to-angular momentum conversion. We found a nonzero orbital OAM in the interference patterns of the far field light obtained from the intensity profile of the outcoupled side waves. The fork fringe pattern was consistent with a vortex of topological charge $|l| \simeq 2$ (Fig. 4b). The intensity measurements of the outcoupled side radiation reported in Fig. 4c were carried out at $\theta_{i}=0°$. We observed a clear modulation in $I(\beta)$ with anticorrelated behaviors between homologous opposite beams, \textit{i.e.} between waves $R$ and $L$, and waves $T$ and $B$. The experimental curves were interpreted with the generation of OAM by considering photon redirection in the momentum space according to a model based on geometric parallel transport of light polarization \cite{bliokh2011,bekshaev2011internal}. The model is fully described in Methods. 

In synthesis, nonparaxial redirection of light due to refraction of partial input plane waves having $\bm{k_{i}} = \frac{2 \pi}{\lambda} \bm{\hat{z}}$ can be described, in the adiabatic approximation, by considering that input polarization is not changed in the local frame attached to the redirected wavevectors $\bm{k}$ \cite{bliokh2011}.  
The output polarization $\bm{{\tilde{E}}_{o}}(\bm{k})$ of a partial wave redirected in the momentum space in the direction $(\theta_{r},\phi)$, \textit{i.e.} having components $k_{x}= k \sin\theta_{r}\cos\phi, k_{y}= k \sin\theta_{r}\sin\phi, k_{z}= k \cos\theta_{r}$, with $k = 2\pi/a$, can be written 
$\bm{{\tilde{E}}_{o}}(\theta_{r},\phi) \propto \hat{U}(\theta_{r},\phi) \bm{E_{i}}$,
where the nondiagonal unitary transformation  $\hat{U}(\theta_{r},\phi)$ is characterized by off-diagonal geometric-phase elements ($\phi$-dependence) that give rise to SOC (Methods). In particular, the key point is that redirection turns a RCP state into a linear combination of RCP and LCP states (Methods). Formally, an input state of OAM $l$ and SAM $\sigma$ is transformed according to $ \ket{l,\sigma} \rightarrow  a \ket{l,\sigma} -b \ket{l+2\sigma,-\sigma} -\sqrt{2ab} \ket{l+\sigma,0}$. Therefore, an input state $\ket{l=0, \sigma}$ is transformed into a combination of opposite helicity states with a nonzero OAM $l = 2\sigma$, so as total angular momentum $l+\sigma = const$. This simple model does not take into account spin-momentum locking arising in structured evanescent waves. Therefore, we weighted the intensities of the outcoupled circular polarization components $|E_{w,j}^{\pm}|^2$ with two phenomenological coefficients $c_{\pm}$ accounting for the efficiency of spin-momentum locking (Methods). In other words, the intensities $I_{j}(\alpha,\beta)$ of the outcoupled waves were measured as a function of the input polarization $\bm{E_{i}}(\alpha,\beta)$ and fitted to 
 \begin{equation}
I_{j}(\alpha,\beta) = c_{+}|E_{w,j}^{+}|^2 + c_{-}|E_{w,j}^{-}|^2 + c_{z}|E_{w,j}^{z}|^2,  \;\; j \in \{R,L,T,B\}.
\end{equation}
In particular, in our case $\theta_{r} = \pi/2$ and $\phi_{T} = 0^{\circ}$, $\phi_{L} = 90^{\circ}$, $\phi_{B} = 180^{\circ}$ and $\phi_{R} = 270^{\circ}$.
This basic model allowed us to interpret all the experimental results here reported (Fig. 4c). Since  the input polarization is turned into a combination of RCP and LCP states, the chiral behavior produces the doubled  periodicity of the curves ($\pi/2$) with a pronounced asymmetric character.   In addition, reversing the input helicity state at a fixed output produces a reversal in the intensity signal behavior as a function of $\beta$, well predicted by our model. In Table 1, we report the fit parameters and the spin-momentum locking efficiency (directivity) $\eta$ (Methods). The fitted coefficient $c_{z}$ was below $1\%$ and thus is not reported. The value $|\eta|$  varied from 32$\%$ to 68$\%$ at $\theta_{i} = 0^{\circ}$.\\
A second set of measurements is reported in Fig. 5.  The capability of tuning the operating wavelength of the resonator structure is an important advantage of this all-dielectric approach, not affected by material losses. This time, data were acquired with a PhCM designed to support a BIC in the infrared, experimentally at 1497 nm. Characterizations were carried out as for the PhCM in the visible. Of course, the laser source was tuned to the corresponding infrared wavelength accordingly. 
Remarkably, $|\eta|$ reached the largest value under a slight deviation from normal incidence, \textit{i.e.} at $\theta_{i} = 0.03^{\circ}$.  As can be seen from Table 2 in Fig. 5, $|\eta|$ reached $94\%$. In addition, the normalized Stokes parameter $s_3$ measured on the outcoupled side beams showed a significant degree of circular polarization, up to $|s_3|=85\%$, regardless of the input polarization state incident on the sample.  The beams on opposite sides of the sample had opposite values of $s_3$. 
 We ascribe this increased spin-controlled coupling to the breaking of the system symmetry due to a non-symmetric excitation condition, which favored the observed chiral coupling (Fig. 5, Table 2). Indeed, the BIC transverse SAM has components also in the PhCM plane (Fig. 2), which may thus influence the overall efficiency of light coupling.
 
Our simple model predicted also the experimental behavior observed upon rotation of the PhCM with respect to the reference system around the $z$ axis (angle $\phi$, Fig. 3a). 
The second row of Fig. 5a shows the effect of rotating by $\phi_0 = 30^{\circ}$ the PhCM. The rotation of the crystal with respect to the input polarizer produced the simple transformation of the geometric phase $\phi_{j}^{'} =\phi_{j}+ \phi_0$ for each redirected beam. The asymmetric shift of the intensity curves is in good agreement with the expected one. (This was repeated for several angles.)  From the fit, we found a remarkable value of directivity $|\eta| = 99.9\%$ with $|s_3| \simeq 1$, as shown in Table 2. 

These results demonstrate a spin-directive coupling due to spin-momentum locking at the BIC frequency in the PhCM, in other words the possibility of a macroscopic quantum spin-Hall effect of light in an all-dielectric nanostructure.
\paragraph*{Conclusions.}
Concluding, light resonant trapping provided by a designer PhC metasurface resonator in the near-BIC regime mediates the coupling to surface waves providing a spatially coherent, well-defined and macroscopic spin-directive coupling of input photons. In analogy to \cite{petersen2014chiral}, the asymmetry observed in the side wave intensities is produced by a strong spin-orbit interaction that breaks the symmetry of the nanoscale system. In our case, the resonant BIC mode enhances the QSHE providing a macroscopic routing of the input photons. Our approach defines a highly efficient and robust far-field excitation scheme for spin-momentum locking in structured optical fields. This mechanism is theoretically interpreted with a simple model based on geometric parallel transport of light polarization in order to fit the experimental data. Since a resonator in the BIC regime  is characterized by strong near-field amplification, engineering such high $Q$-factor resonators may lead to metasurface multiplatform applications, also favored by the large scale collective character of the BIC that makes the resonator intrinsically robust against external perturbations. Concurrently, other effects may be enhanced, such as unidirectional low-threshold lasing and strong-coupling regime phenomena of nanoscale supercavities, providing novel quantum chiral photonic applications.

% Your references go at the end of the main text, and before the
% figures.  For this document we've used BibTeX, the .bib file
% scibib.bib, and the .bst file Science.bst.  The package scicite.sty
% was included to format the reference numbers according to *Science*
% style.

%\bibliography{Myscibib}

\bibliographystyle{Science}

\newpage
%%FIGURE 1
\begin{figure}[h!]
\includegraphics[width=12 cm]{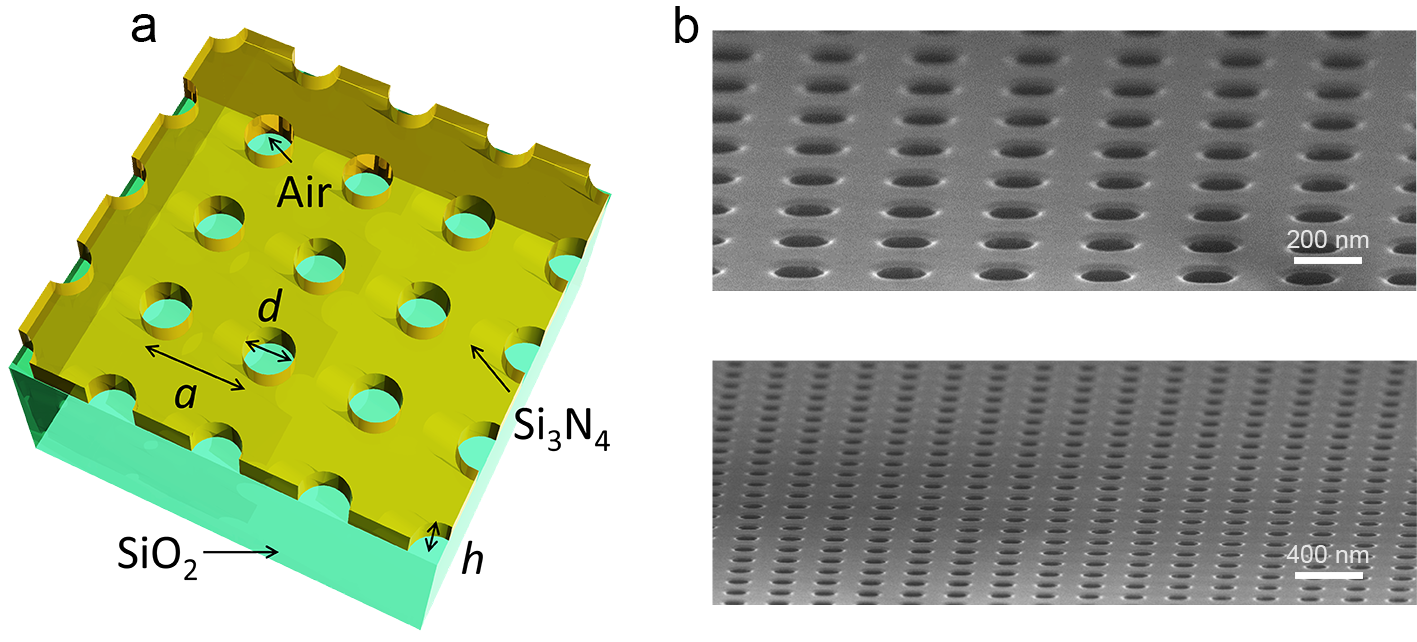}
\centering
\caption{\label{fig_big}   \textbf{a.} Schematic layout of the PhCM. The experimental sample consists of square lattice of period $a$ of cylindrical air holes etched in a silicon nitride (Si$_3$N$_4$) thin film ($h \simeq \lambda /10 = 54$ nm). The silicon nitride film covers all the surface of the supporting quartz substrate (SiO$_2$), which has thickness 120 $\mu$m. The patterned area is 1 mm$^2$. \textbf{b.} Scanning electron microscopy images of the patterned area (tilted view). }
\end{figure}

%%FIGURE 2
\newpage
\begin{figure}[h!]
\includegraphics[width=15 cm]{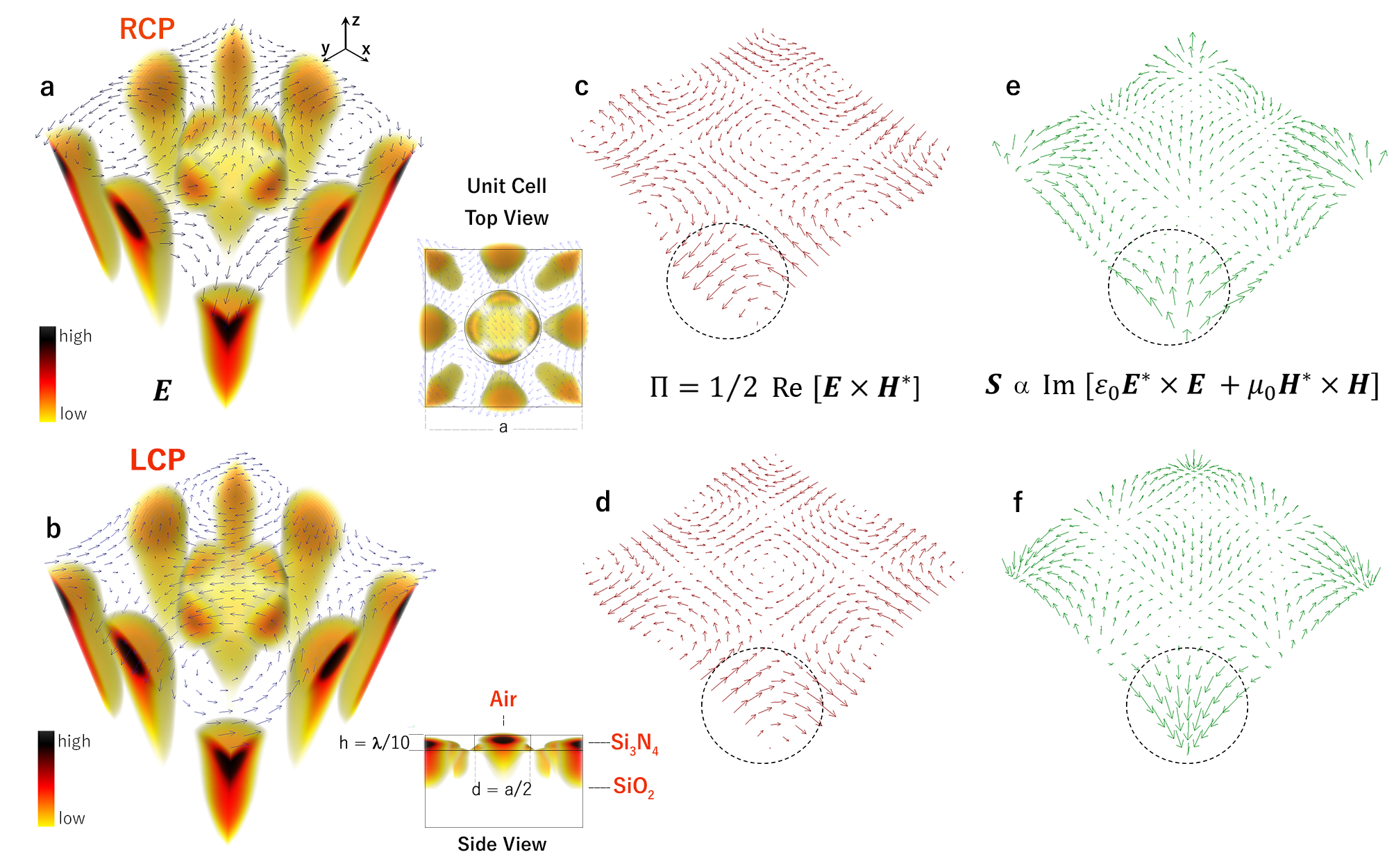}
\centering
\caption{\label{sim} \textbf{FDTD numerical analysis of the electromagnetic field at the BIC and input SAM dependence.} The top row refers to results obtained for RCP, whereas the second row to LCP excitation. \textbf{a-b.} $|\bm{E}|^2$-map inside the unit cell of the square lattice in case of input RCP (3$D$ view according to reference system, values below a threshold intensity are masked for a better visualization) together with the vector map of $\bm{E}$ at $z = h/2$ of the PhC. The electric field is structured in a vortex-antivortex lattice and shows cycloidal rotation in the planes orthogonal to the $(x, y)$-plane. The insets show a top view and a side view of the unit cell of the PhCM. The side view of the intensity of the BIC shows its surface wave character. \textbf{c-d.} The Poynting vector maps at the BIC clearly shows that light is confined in the $(x, y)$-plane of the crystal, and photon flow is reversed upon spin flip of the incident excitation radiation. 
\textbf{e-f.} The  SAM density maps show a pronounced component that is transverse to the the $(x, y)$-plane and transverse to the direction of propagation. This last is locked to the sign of the spin which is determined by the input spin value of the exciting radiation, as visible in the dashed circles comparing $\bm{\Pi}$ and $\bm{S}$ for both RCP and LCP.}
\end{figure}
\newpage

%Figure 3
\begin{figure}[h!]
\includegraphics[width=16 cm]{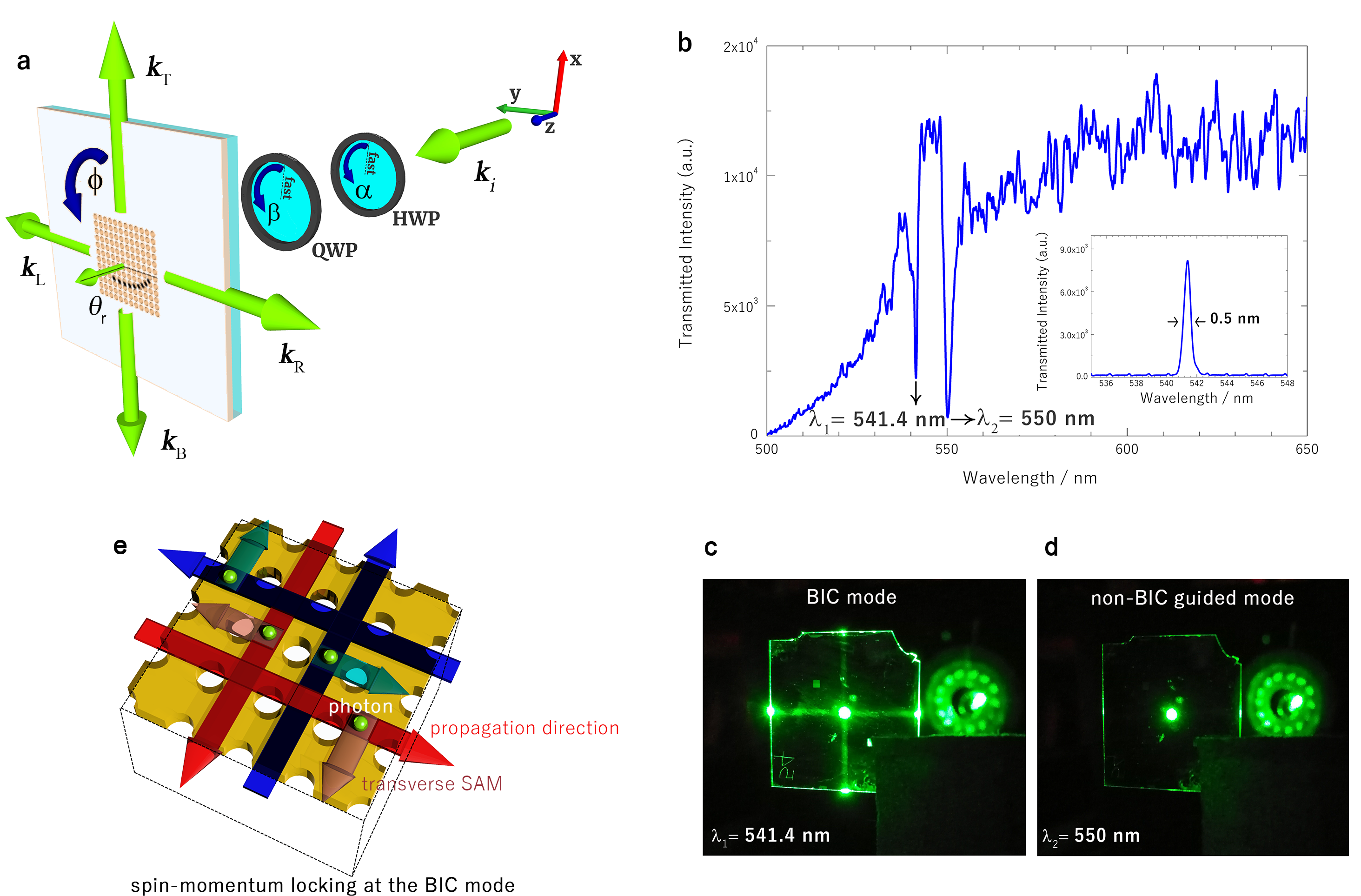}
\centering
\caption{\label{fig_big} \textbf{Phenomenology of the redirection effect at the BIC.}  \textbf{a.} Schematic layout of the experimental setup. The input beam polarization $\bm{E_i}$ (initially $\parallel \bm{\hat{x}}$) is controlled by means of an half wave plate (HWP) and a quarter wave plate (QWP). At the BIC frequency, light is redirected at $\theta_r = \pi/2$ along the PhC axes of symmetry.  \textbf{b.} Normal incidence transmission spectrum of a PhCM designed to support a BIC in the visible range. Two modes are visible: $\lambda_1$ corresponds to a near-BIC regime, whereas $\lambda_2$ to a conventional leaky mode. \textbf{c.} At the BIC wavelength $\lambda_1$, the normally incident light is experimentally redirected along the PhC symmetry axes. The inset of the spectrum shows the far field spectral profile of the redirected beam of wavevector $\bm{k}_{R}$ collected from the side corner of the sample. \textbf{d.} At $\lambda_2$ there is no detectable redirection effect.
 \textbf{e.}  The cartoon shows the schematics of the principle of spin-momentum locking due to the transverse SAM of the evanescent waves at the BIC in the PhCM (in the cartoon only planar components are depicted for simplicity). Since light confinement at the BIC generates evanescent waves characterized by transverse SAM, photons at the boundaries of the metasurface have momenta locked to their transverse spin and orientation of the imaginary component of the wavevector. The spin value of the incident radiation determines the sign of the transverse SAM, hence the sign of the wavevectors. }
\end{figure}

\newpage

\begin{figure}[h!]
\includegraphics[width=16 cm]{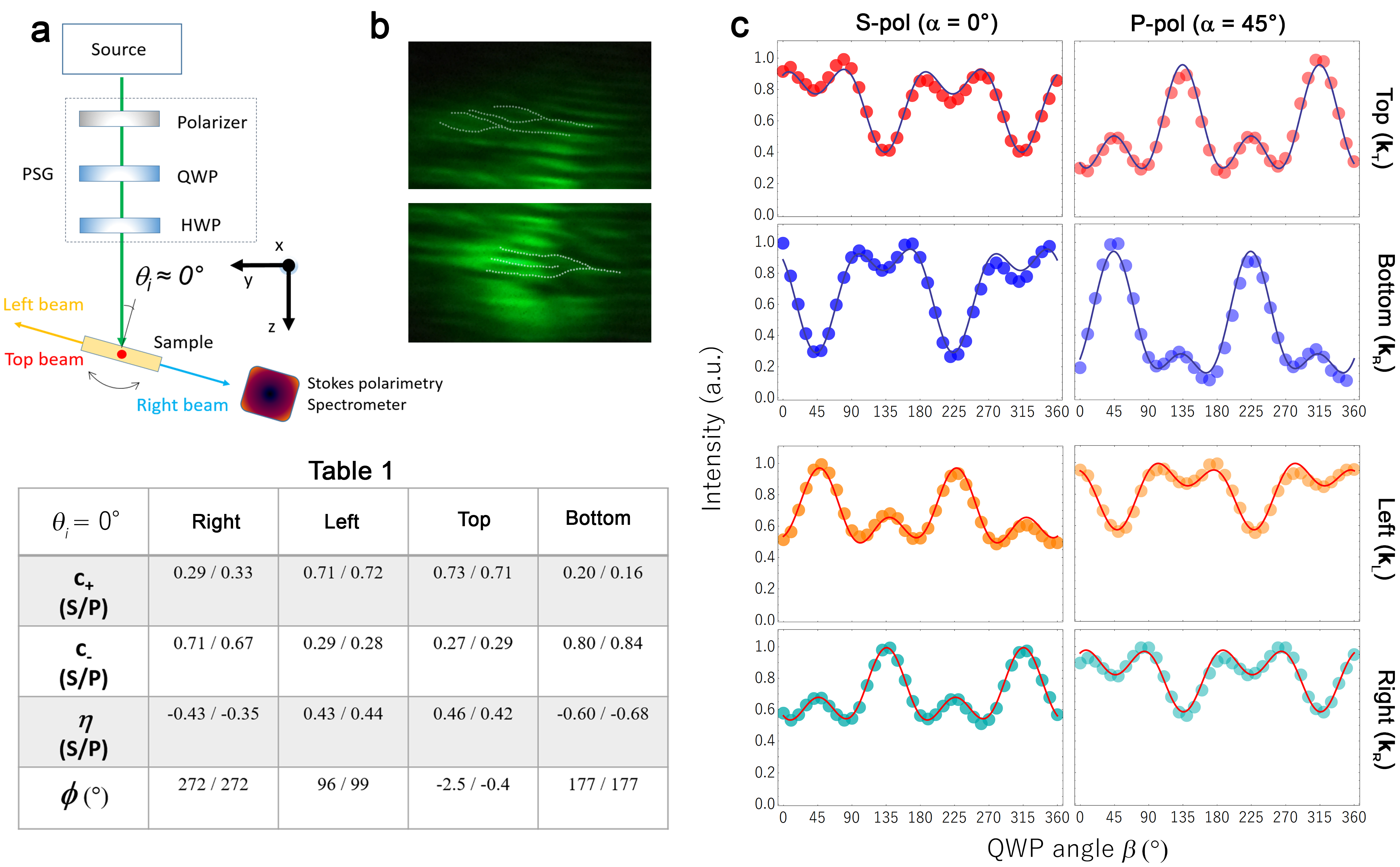}
\centering
\caption{\textbf{Chiral behavior at the BIC (normal incidence): experimental results and fit based on parallel transport of polarization and spin-momentum  locking}. \textbf{a.} Layout of the far field characterization: measurements were carried out at normal incidence, $\theta_i = 0°$, with the PhCM axes parallel to the laboratory reference system, as indicated in Fig. 1a. \textbf{b.} The dislocations observed in the far field interference pattern of a side beam are consistent with the generation of an OAM $|l|$ = 2.  \textbf{c.} Intensity of the redirected beams outcoupled from the PhCM (rows) measured as function of the QWP angle $\beta$  for input S- and P-polarization incident on the QWP (columns).  The experimental points show a chiral behavior that can be explained with the generation of opposite spin photons due to SOC and spin-momentum locking, as shown by the excellent agreement to the model (solid lines) in which spin-momentum locking is taken into account by means of phenomenological chiral coefficients. The chiral parameters resulting from the fit are indicated in \textbf{Table 1}. }
\end{figure}
\newpage
\begin{figure}[h!]
\includegraphics[width=17 cm]{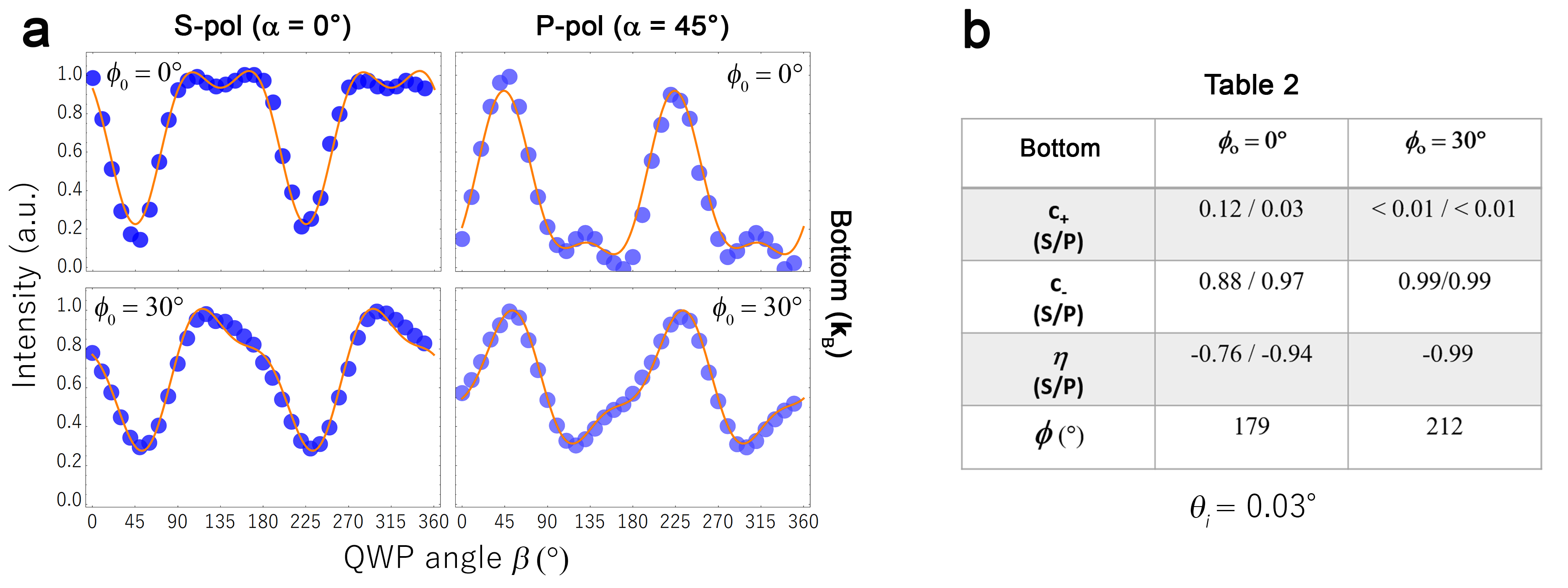}
\centering
\caption{\textbf{Chiral behavior with broken symmetry excitation and geometric phase effect}. \textbf{a.} Intensity of the redirected beams outcoupled from the PhCM (rows) measured as function of the QWP angle $\beta$  for input S- and P-polarization incident on the QWP (columns). This time, measurements were carried out at $\theta_i = 0.03^{\circ}$ with the PhCM axes parallel to the laboratory reference system, $\phi_0 = 0^{\circ}$ (first row), and after a rotation of the PhCM to $\phi_0 = 30^{\circ}$ (second row). 
\textbf{b.} The experimental points show chiral behaviors in accordance with SOC prediction. Remarkably, in this case, spin-momentum locking efficiency $\eta$ is close to 1, as evident from the fit parameters in \textbf{Table 2}.}
\end{figure}

\newpage

\section*{Supplementary Materials}
\subsection*{Methods}

\textbf{Sample fabrication}. The PhCM samples were 1 $\times$1 mm$^2$ large  and consisted of air-cylindrical holes in a square lattice. The patterned area was realized on a Si$_{3}$N$_{4}$ layer deposited on a quartz slide by means of plasma-enhanced chemical vapor deposition (PECVD).  The square lattice was defined on ZEP 520, a positive electron beam resist, by using a high-precision nanofabrication process based on high-voltage electron beam lithography  (Vistec VB300UHR EWF). The resist mask was then transferred by reactive ion etching (RIE) in an Oxford Instruments PlasmaLab 80 Plus tool, using CHF$_{3}$ and O$_{2}$, at room temperature. Figure 1b shows the final result. The lattice parameters were optimized in order to have the resonant behavior of the structure around 540 nm for the sample in the visible range and 1550 nm for the sample in the infrared range. The lattice hole radius was  $r = a/4$, whereas the thickness of the PhCM was $h = 0.1 \lambda$, with $\lambda$ given by the free-space wavelength of the expected BIC. Such values were determined according to the procedure described in Ref. \cite{mocella2015giant}.    

\textbf{Simulations}. Numerical simulation of the excited modes in the structure were carried out by using a full three-dimensional rigorous coupled wave approach (RCWA) based on a Fourier modal expansion.
Additional finite difference time domain (FDTD) simulations,  using commercial software FULLWAVE (https://optics.synopsys.com/rsoft/rsoft-passive-device-fullwave.html) allowed us eliciting the spin-dependent character of the the excited states. The computational domain was limited to a unit cell of the PhCM.  We applied Bloch periodic boundary condition to surfaces along $x$-, $y$-directions. On top and bottom surfaces, normal to the $z$-direction, and far enough from the membrane, we imposed perfectly-matched-layer absorbing boundary conditions \cite{zito2016dark,zito2016en}. The adapted mesh along $z$ had a size-step of 3 nm inside the PhCM and increased outside, up to a value of 20 nm. The time step was chosen to be well below the stability limit. Steady-state regime was reached after 10$^5$ optical cycles. As a further check, finite element method-based simulations, carried out with Comsol Multiphysics 5.2a, were finally used to verify the consistence of all simulations and the properties of the symmetry-protected BIC modes.
 Once determined the electromagnetic field $(\bm{E},\bm{H})$ of the modes, we calculated the linear momentum density $\bm{\Pi}$ and SAM density $\bm{S}$ according to Ref. \cite{bliokh2015transverse,aiello2015}, as described in the main text. The amplification of the field is limited only by the numerical precision of the simulation, reaching values as large as $10^8 E_i$, with $E_i$ input field amplitude. 

\textbf{Measurements}. Samples were investigated by means of a supercontinuum laser (SuperK Extreme NKT Photonics), which allows a high output power, a broad spectrum, and a high degree of spatial coherence. The emission spectrum was 400 - 2400 nm. By using acousto-optic tunable filters, we converted it into an ultra-tunable laser with up to 8 simultaneous laser lines.
The sample was mounted on a motorized rotation stage, allowing rotational steps of $0.01^\circ$. The transmitted spectrum was collected by a spectrometer (HR4000, Ocean Optics) with a resolution of 0.25 nm and a range of 350 - 800 nm. The collimated beam incident on the sample had a beam waist of $\sim 0.5$ mm. The radiation outcoupled from the sides of the sample was analyzed to determine the Stokes parameters of each beam, in order to measure the degree of circular polarization.   \\
In an ideal PhCM, the quality factor of a BIC peak tends to infinity. However, in real devices, the resonance
width is limited by finite sample size, effective collimation and scattering losses due to the sample imperfections \cite{gansch2016measurement}. As such, the total $Q$-factor of the resonator
can be written as $1/Q = 1/Q_{V} + 1/Q_{H} +1/ Q_{A}$, in which 
the term $Q_{V}$ depends on the coupling efficiency to incident light and corresponds to the intrinsic BIC quality factor;  $Q_{H}$ measures the horizontal
in-plane losses due to the finite lateral extension of the pattern; and 
$Q_{A}$ accounts for the material absorption and scattering losses. The measured $Q$-factor is consistent with a near-BIC regime, as corresponding to a linewidth $\simeq$ 0.5 nm (Fig. 3b) due to finite lateral confinement as in previous reports \cite{gansch2016measurement,hsu2013}. 

\textbf{SOC model based on Geometric Parallel Transport of Polarization}. The coherent scattering process of a photon propagating along $\bm{\hat{z}}$ and redirected along $\pm \bm{\hat{x}}$ and $\pm \bm{\hat{y}}$ in the plane of the PhC can be described using the theory of geometric parallel transport of light polarization. We followed the framework described in \cite{bliokh2011,bliokh2015,bliokh2015transverse}, which we explicitly refer to in the following. 
Let us consider a paraxial optical field of polarization $\bm{E_{i}}(\bm{r}) \equiv (E^{+}, E^{-}, 0)^{T}$, \textit{i.e.} of circular components $E^{+}$ and $E^{-}$ and null longitudinal $z$-component in the global circular representation of polarization. 
Circular basis vectors are related to the Cartesian linear representation through  $\bm{u_{\pm}} = \frac{1}{\sqrt{2}}(\bm{u_{x}}\pm i \bm{u_{y}})$, whereas the components can be transformed according to $E^{\pm} = \frac{1}{\sqrt{2}}(E_{x}\mp i E_{y})$. Given respectively $\alpha$ and $\beta$ the orientation angles of the HWP and QWP fast axes with respect to the input $\bm{u_x}$-polarizer (Fig. 1a), 
their Jones matrices $\hat{H}(\alpha)$ and $\hat{Q}(\beta)$ are expressed in the linear representation, for an ideal paraxial field with null $z$-component, according to
\begin{equation}
 \hat{H}(\alpha)  =\left(
{\begin{array}{ccc}
 \cos (2 \alpha) & \sin (2 \alpha) &0\\
 \sin (2 \alpha) & -\cos (2 \alpha) &0\\
 0&0&0\\
\end{array}}
\right),
 \end{equation}
 
 \begin{equation}
 \hat{Q}(\beta)  =\left(
\begin{array}{ccc}
 \cos ^2(\beta)+i \sin ^2(\beta) & (1-i) \sin (\beta) \cos (\beta)&0 \\
 (1-i) \sin (\beta) \cos(\beta) & i \cos ^2(\beta)+\sin ^2(\beta) &0\\
0&0&0\\
\end{array}
\right).
 \end{equation}
Therefore, the input paraxial field on the PhCM can be written as
 \begin{equation}
 \bm{E_{i}}(\alpha,\beta) = \hat{V_c}\hat{Q}(\beta)\hat{H}(\alpha)E_{x}\bm{u_{x}},
 \end{equation}
 being $E_{x}\bm{u_{x}}$ the input laser state and $\hat{V_c}$ the Cartesian-to-circular representation transform
 \begin{equation}
 \hat{V_c} =\frac{1}{\sqrt{2}}\left(
\begin{array}{ccc}
1 &i&0\\
 1&-i&0\\
0&0&\sqrt{2}\\
\end{array}
\right).
 \end{equation}
Non-paraxial redirection of light due to refraction of partial input plane waves having $\bm{k_{i}} = \frac{2 \pi}{\lambda} \bm{\hat{z}}$ can be described, in the adiabatic approximation, by considering that input polarization is not changed in the local frame attached to the redirected wavevectors $\bm{k}$ \cite{bliokh2011}.  
The output polarization $\bm{{\tilde{E}}_{o}}(\bm{k})$ of a partial wave redirected in the momentum space in the direction $(\theta_{r},\phi)$, \textit{i.e.} having components $k_{x}= k \sin\theta_{r}\cos\phi, k_{y}= k \sin\theta_{r}\sin\phi, k_{z}= k \cos\theta_{r}$, with $k = 2\pi/a$, can be written 
\begin{equation}
\bm{{\tilde{E}}_{o}}(\theta_{r},\phi) \propto \hat{U}(\theta_{r},\phi) \bm{E_{i}}. 
\end{equation}
The angles $(\theta_{r},\phi)$
serve both as real-space coordinates for the incident field $\bm{E_{i}}$ (for which $\theta_{r} = \theta_{i} = 0$) and momentum-space
coordinates for the refracted field $\bm{{\tilde{E}}_{o}}$.
The nondiagonal unitary transformation  $\hat{U}(\theta_{r},\phi)$ is defined in the global circular basis as
\begin{equation}
 \hat{U}(\theta_{r},\phi)  =\left(
 {\begin{array}{ccc}
   a & -b e^{-2 i \phi} &\sqrt{2 a b } e^{-i \phi}\\
   -b e^{2 i \phi} & a & \sqrt{2 a b } e^{i \phi}\\
   -\sqrt{2 a b } e^{i \phi}& -\sqrt{2 a b } e^{-i \phi} & a-b\\
  \end{array} }
\right),
\end{equation}
with $a=\cos^2{\theta_{r}/2}$ and $b = \sin^2{\theta_{r}/2}$, and it is characterized by off-diagonal geometric-phase elements ($\phi$-dependence) from which \textit{spin-to-orbital angular momentum conversion} stems. The angle $\theta_{r} $ belongs to the range $(0, \theta_{c})$, where in the limiting case $\theta_{c} = \pi/2$, whereas $\phi \in (0, 2\pi)$. 
The SOC can be generalized formally by considering an input state having helicity $\sigma$ and orbital angular momentum per photon $l$ with respect to the propagation direction $\bm{\hat{z}}$. The input circularly polarized vortex of helicity state $\sigma$ and OAM $l$ will be transformed in a new state characterized by the emergence of a component of opposite helicity $-\sigma$ and change in the orbital angular momentum $l + 2 \sigma$. Symbolically, this can be written as 
the transformation  
\begin{equation}
 \ket{l,\sigma} \rightarrow  a \ket{l,\sigma} -b \ket{l+2\sigma,-\sigma} -\sqrt{2ab} \ket{l+\sigma,0}. 
\end{equation}
The case of zero input orbital angular momentum $\ket{l = 0,\sigma} $  is a special case. Nonetheless, in such a case, the output field assumes a defined orbital angular momentum $l = 2 \sigma$, obeying conservation of total angular momentum $l+s = const$. 

Let us consider the four redirected beams with wavevectors $\textbf{k}_{R} = -k \bm{\hat{y}}, \textbf{k}_{L}=+k\bm{\hat{y}}, \textbf{k}_{T}=+k\bm{\hat{x}},\textbf{k}_{B}=-k\bm{\hat{x}}$ in our experimental case (Fig. 3a). These are produced by the 
PhCM at the BIC frequency. The overall effect of the BIC mode excitation, formally, is in that the incoming plane wave is abruptly refracted by the photonic structure in the momentum space along 4 specific directions dictated by the crystal symmetry. The fields  of the redirected waves $\bm{{\tilde{E}}_{o}}({\textbf{k}}_{j})$, with $j = \{R,L,T,B\}$,   combine themselves with the incident field over the PhC volume. We conjectured that the experimentally outcoupled waves of polarization $\bm{{\tilde{E}}_{w}^{(j)}}({\textbf{k}}_{j})$ at angles $\theta_{j} = \theta_c = \frac{\pi}{2}$,  with $\phi_{j} = (-\frac{\pi}{2}, \frac{\pi}{2}, 0, \pi)$, respectively, were proportional to the total field  $\bm{{\tilde{E}}_{o}}({\textbf{k}}_{j})+ \bm{{E}_{i}}$. The electric fields $\bm{{\tilde{E}}_{o}}({\textbf{k}}_{j})$ are obtained according to the geometric parallel transport transformation mediated by the operator $\hat{U}(\theta_{r},\phi) $. Finally, this reads
\begin{equation}
\bm{{\tilde{E}}_{w}^{(j)}}\left(\frac{\pi}{2},\phi_{j}\right) \equiv (E_{w,j}^{+}, E_{w,j}^{-}, E_{w,j}^{z})^{T} \propto \left[\hat{U}\left(\frac{\pi}{2},\phi_{j}\right) + \mathcal{\hat{I}}\right] \bm{E_{i}},
\end{equation}
where $\mathcal{\hat{I}} = {\rm diag}(1,1,1)$.
 This simple model does not take into account spin-momentum locking arising in structured evanescent waves. Indeed, as supported by our numerical simulations predicting spin-dependent directive coupling (Fig. 2), the redirected outcoupled waves are generated by inhomogeneous optical fields characterized by transverse SAM densities. Therefore, we weighted the intensities of the outcoupled polarization components $|E_{w,j}^{\pm}|^2$ with two phenomenological coefficients $c_{\pm}$ accounting for the efficiency of spin-momentum locking. In other words, the intensities $I_{j}(\alpha,\beta)$ of the outcoupled waves were measured as a function of the input polarization $\bm{E_{i}}(\alpha,\beta)$ and fitted to 
 \begin{equation}
I_{j}(\alpha,\beta) = c_{+}|E_{w,j}^{+}|^2 + c_{-}|E_{w,j}^{-}|^2 + c_{z}|E_{w,j}^{z}|^2,  \;\; j \in \{R,L,T,B\}.
\end{equation}
In particular, we fixed the HWP angle $\alpha$ to two values,  $\alpha^{S} = 0$ ($S$-polarized input on the QWP) and $\alpha^{P} = \pi/4$ ($P$-polarized input on the QWP), and measured the intensities as a function of the QWP angle $\beta \in (0, 2\pi)$.
This basic model allowed us to interpret all the experimental results here reported and to predict the experimental behavior actually observed upon rotation of the PhCM about the $z$ axis with respect to the reference system. Indeed, the rotation of the crystal produced a variation of the geometric phase $\phi_{j}$ (for each redirected beam) with respect to the global reference system (input polarizer). The fitted chiral ratio
 \begin{equation}
\eta = \frac{c_{+}-c_{-}}{c_{+}+c_{-}},
\end{equation} 
 reached a value $\sim$ 99.9$\%$ under optimal optical coupling, \textit{i.e.}  breaking the symmetry of the excitation at an angle of incidence $\theta_{i} = 0.03^\circ$. In addition, opposite homologous waves, that is along $\mp \bm{\hat{y}}$ (R, L) and $\pm \bm{\hat{x}}$ (T, B), were characterized by opposite sign of $\eta$ (see Table 1,2 in Fig. 4,5).

\newpage

\begin{figure}[h!]
\includegraphics[width=12 cm]{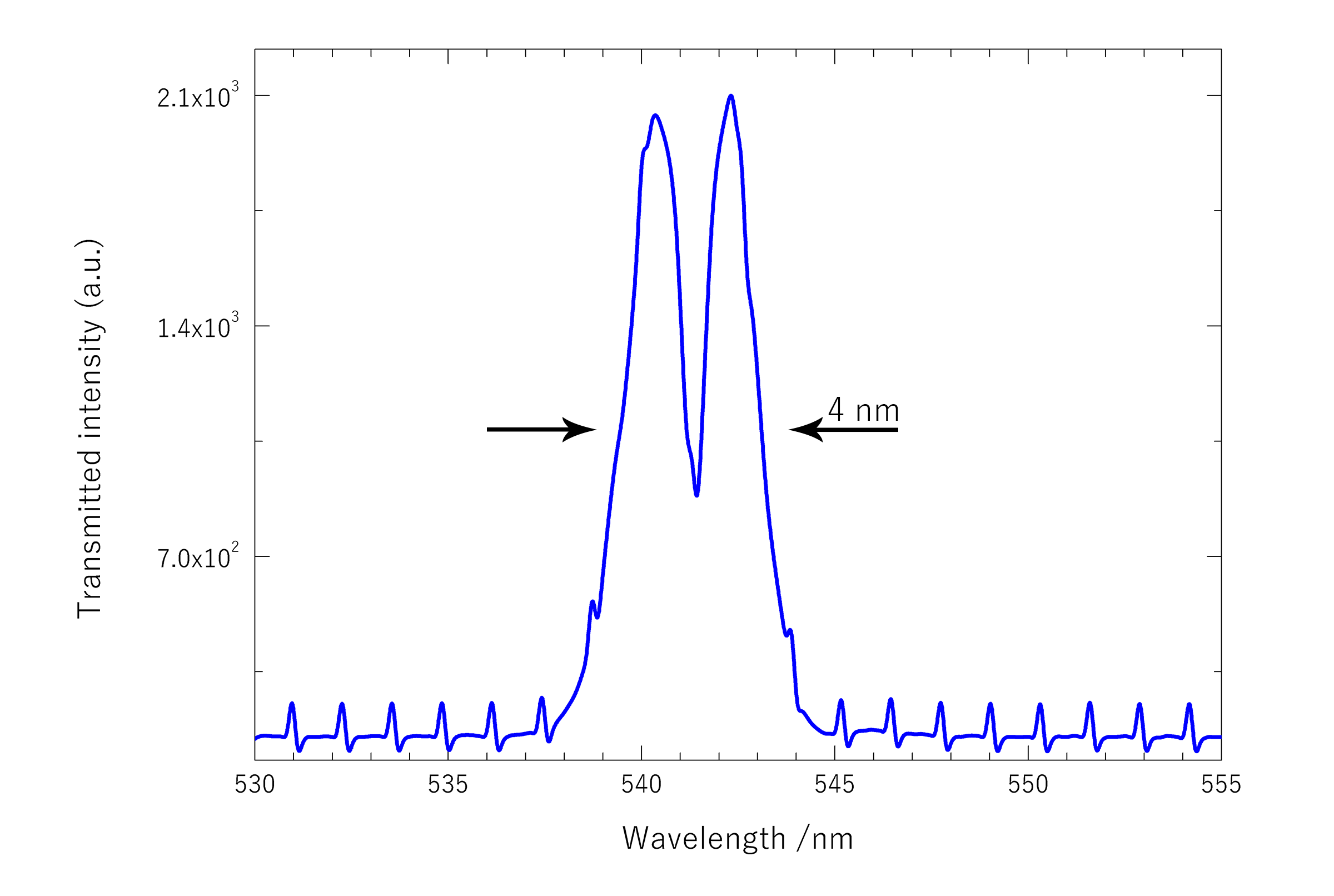}
\centering
\caption{\textbf{Supplementary Fig S1. Transmission spectrum through the PhCM excited with a single laser line centered at 541 nm.} The supercontinuum laser was equipped with acousto-optic tunable filters allowing tunable filtering in the visible and near infrared. The power of each wavelength was controlled via software. The line bandwidth depends on the selected range. When the wavelength was set at 541 nm, a resonance peak appeared overlapped to the laser line (bandwidth 4 nm).}
\end{figure}

\begin{figure}[h!]
\includegraphics[width=12 cm]{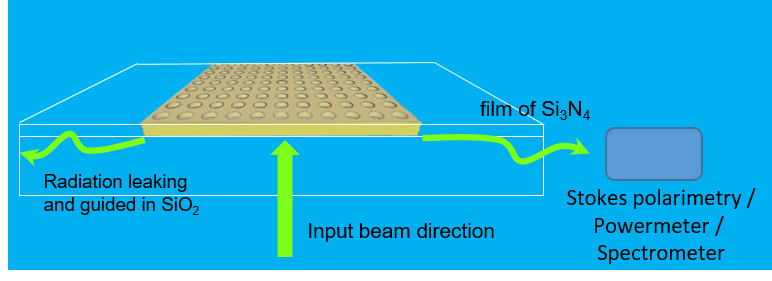}
\centering
\caption{\textbf{Supplementary Fig S2.} Cartoon depicting the process of radiation leaks guided in SiO$_2$.}
\end{figure}

\end{document}